%% file: main.tex
\documentclass[sigconf,nonacm]{acmart}

\usepackage{subcaption}
\usepackage{cleveref}
\usepackage{multirow}
\usepackage{booktabs} %
\usepackage{multirow} %
\usepackage{subcaption}
\usepackage[breakable, skins]{tcolorbox}
\usepackage{adjustbox}
\usepackage{booktabs}
\usepackage{tabularx}   %
\usepackage{soul}
\usepackage{xcolor}
\usepackage{gensymb}
\usepackage{colortbl}
\usepackage{textcomp}
\usepackage{siunitx}

\colorlet{LightOrange}{white!50!orange}
\colorlet{LightBlue}{white!60!blue}
\colorlet{LightGreen}{white!60!green}
\colorlet{LightRed}{white!60!red}

\tcbset{arc=0mm,size=fbox,attach title to upper={\ ---\ },coltitle=black}

\crefformat{section}{\S#2#1#3}
\crefformat{subsection}{\S#2#1#3}
\crefformat{subsubsection}{\S#2#1#3}
\crefrangeformat{section}{\S\S#3#1#4 to~#5#2#6}
\crefmultiformat{section}{\S\S#2#1#3}{ and~#2#1#3}{, #2#1#3}{ and~#2#1#3}

\AtBeginDocument{%
  }

\begin{document}

\title{Investigating Web Content Delivery Performance over Starlink}

\author{Rohan Bose}
\affiliation{%
   \institution{Technical University of Munich}
   \country{Germany}}

\author{Jinwei Zhao}
\affiliation{%
   \institution{University of Victoria}
   \country{Canada}}

\author{Tanya Shreedhar}
\affiliation{%
   \institution{Delft University of Technology}
   \country{Netherlands}}

\author{Jianping Pan}
\affiliation{%
   \institution{University of Victoria}
   \country{Canada}}

\author{Nitinder Mohan}
\affiliation{%
   \institution{Delft University of Technology}
   \country{Netherlands}}

\input{sections/abstract.tex}

\maketitle

\vspace*{-1em}
\input{sections/intro.tex}

\input{sections/methodology.tex}

\input{sections/global.tex}

\input{sections/africa}

\input{sections/related.tex}
\vspace*{-1em}
\input{sections/discussion}

\bibliographystyle{ACM-Reference-Format}
\bibliography{references}

\balance
\appendix
\input{sections/app-ethics.tex}

\input{sections/app-measurement_config}
\input{sections/app-additional_analysis}

\end{document}

%% file: sections/abstract.tex
\begin{abstract}

Low Earth Orbit (LEO) satellite ISPs promise universal Internet connectivity, yet their interaction with content delivery remains poorly understood.
We present the first comprehensive measurement study decomposing Starlink's web content delivery performance decomposed across Point of Presence (PoP), DNS, and CDN layers.
Through two years of measurements combining 225~K Cloudflare AIM tests, M-Lab data, and active probing from 99 RIPE Atlas and controlled Starlink probes, we collect 6.1~M traceroutes and 10.8~M DNS queries to quantify how satellite architecture disrupts terrestrial CDN assumptions.
We identify three distinct performance regimes based on infrastructure density.
Regions with local content-rich PoPs achieve near-terrestrial latencies with the satellite segment dominating 80--90\% of RTT.
Infrastructure-sparse regions suffer cascading penalties: remote PoPs force distant resolver selection, which triggers CDN mislocalization, pushing latencies beyond 200\,ms.
Dense-infrastructure regions show minimal sensitivity to PoP changes.
Leveraging Starlink's infrastructure expansion in early 2025 as a natural experiment, we demonstrate that relocating PoPs closer to user location reduces median page-fetch times by 60\%.
Our findings reveal that infrastructure proximity, \emph{not satellite coverage}, influences web performance, requiring fundamental changes to CDN mapping and DNS resolution for satellite ISPs.

\end{abstract}

%% file: sections/intro.tex
\section{Introduction}

Over the past three decades, Internet infrastructure, technologies and use cases have evolved from interconnecting computers on a global network to a platform for content distribution.
Today, roughly 5.5 billion people generate nearly 330 exabytes of traffic daily, with content delivery networks (CDNs) shouldering close to 70\% of this load~\cite{CiscoAnnualReport,sandvine_gipr_2024}.
Simultaneously, the majority of CDN-driven applications are latency-sensitive, including video streaming, online gaming, and web browsing, driving the requirement for user requests to be served by the ``closest'' CDN cache server with minimal latency~\cite{cisco-vni-2019}.
To meet this demand cost-effectively, CDN operators deploy thousands of cache servers within major Internet service provider (ISP) networks and at metropolitan Internet exchange points (IXPs), strategically placing these servers both topologically and geographically \textit{near} large user populations~\cite{akamai-sigops2010,wohlfartLeveragingInterconnectionsPerformance2018,gigisSevenYearsLife2021}.
CDN providers employ various request-mapping techniques, well-studied in the literature, to systematically steer traffic to the ``closest'' cache server.
The most widely-deployed techniques, including IP anycast~\cite{calder2015analyzing}, DNS-based geolocation~\cite{akamai-sigops2010}, and dynamic request routing, leverage the inherent Internet topology to identify the optimal and geographically nearest cache server for each user~\cite{krishnan2009moving,zhu2012latlong,odin-nsdi18,valincus2013quantifying,endusermapping-sigcomm15,schlinkerEngineeringEgressEdge2017}.
Effective request-mapping techniques are critical for minimizing cache misses, which would otherwise lead to traversals to remote origin servers, resulting in inflated latency and degraded Quality of Experience (QoE)~\cite{zhou2023regional}.

\begin{figure}[t!]
    \centering
    \includegraphics[width=\linewidth]{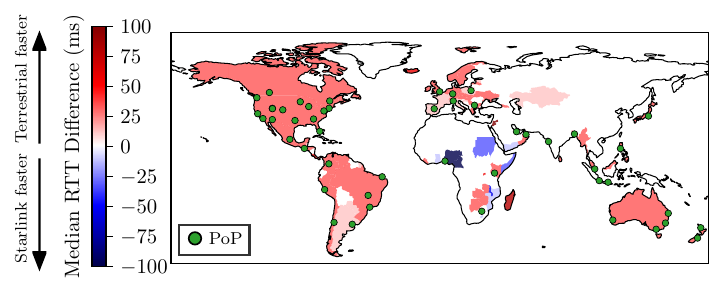}
    \vspace*{-2.5em}
    \caption{Median RTT difference (Starlink $-$ Terrestrial) between Starlink subscribers and Terrestrial subscribers to the best Cloudflare CDN server reported by Cloudflare AIM.}
    \label{fig:starlink-terrestrial-comparison}
    \Description{Starlink vs Terrestrial RTT comparison}
    \vspace*{-1.5em}
\end{figure}

In the last few years, LEO satellite networks (LSNs) have emerged as a new access technology, providing Internet connectivity throughout the globe~\cite{multifacetedStarlink-www24, suizoFirstLookStarlinks2025, wang2024large}.
LSNs, spearheaded by SpaceX's Starlink~\cite{starlink, starlink-coverage} alongside competitors such as Eutelsat's OneWeb~\cite{oneweb}, are evolving into ``global ISPs'' that challenge traditional Internet operational models~\cite{osoroTechnoEconomicFrameworkSatellite2021}, bridge connectivity gaps in remote regions~\cite{mcmahonAssessingImpactsLowearth2025,suizoFirstLookStarlinks2025}, provide service in offshore and mobile environments~\cite{laniewskiMeasuringMobileStarlink2025a,liFundamentalsSatelliteMaritimeCommunications2025}, and offer competitive alternatives to existing terrestrial networks (TNs)\footnote{We refer to terrestrial ISPs as ``TNs'' and LEO ISPs as ``LSNs'' throughout the paper.} in well-provisioned regions~\cite{huLEOSatelliteVs2023b}.
The research community has devoted significant attention to understanding LSN end-to-end performance~\cite{liz2024netflix,zhao2023real}, unique traffic engineering challenges~\cite{weiEfficientTrafficEngineering2024,wuSaTELowLatencyTraffic2025}, and impact on Internet routing~\cite{handleyDelayNotOption2018a,pan2024measuring}, with recent studies demonstrating that LSNs can match or exceed TN performance in latency and throughput~\cite{multifacetedStarlink-www24}.
However, we find that LSNs challenge fundamental architectural assumptions of the traditionally terrestrial Internet, resulting in significant performance overheads, the extent of which has received limited attention within the community.
\Cref{fig:starlink-terrestrial-comparison} illustrates a preliminary comparison of latency to well-known websites from Starlink versus cellular terrestrial connections from the same country (methodology described in \Cref{sec:methodology:passive}).
We observe that Starlink apart from African subcontinent, Starlinks lags behind TNs in web performance almost globally; however, we find that CDN infrastructure proximity to PoP, \emph{not satellite coverage} critically impacts end-to-end performance.

\begin{figure}[t!]
    \centering
    \includegraphics[width=0.7\linewidth]{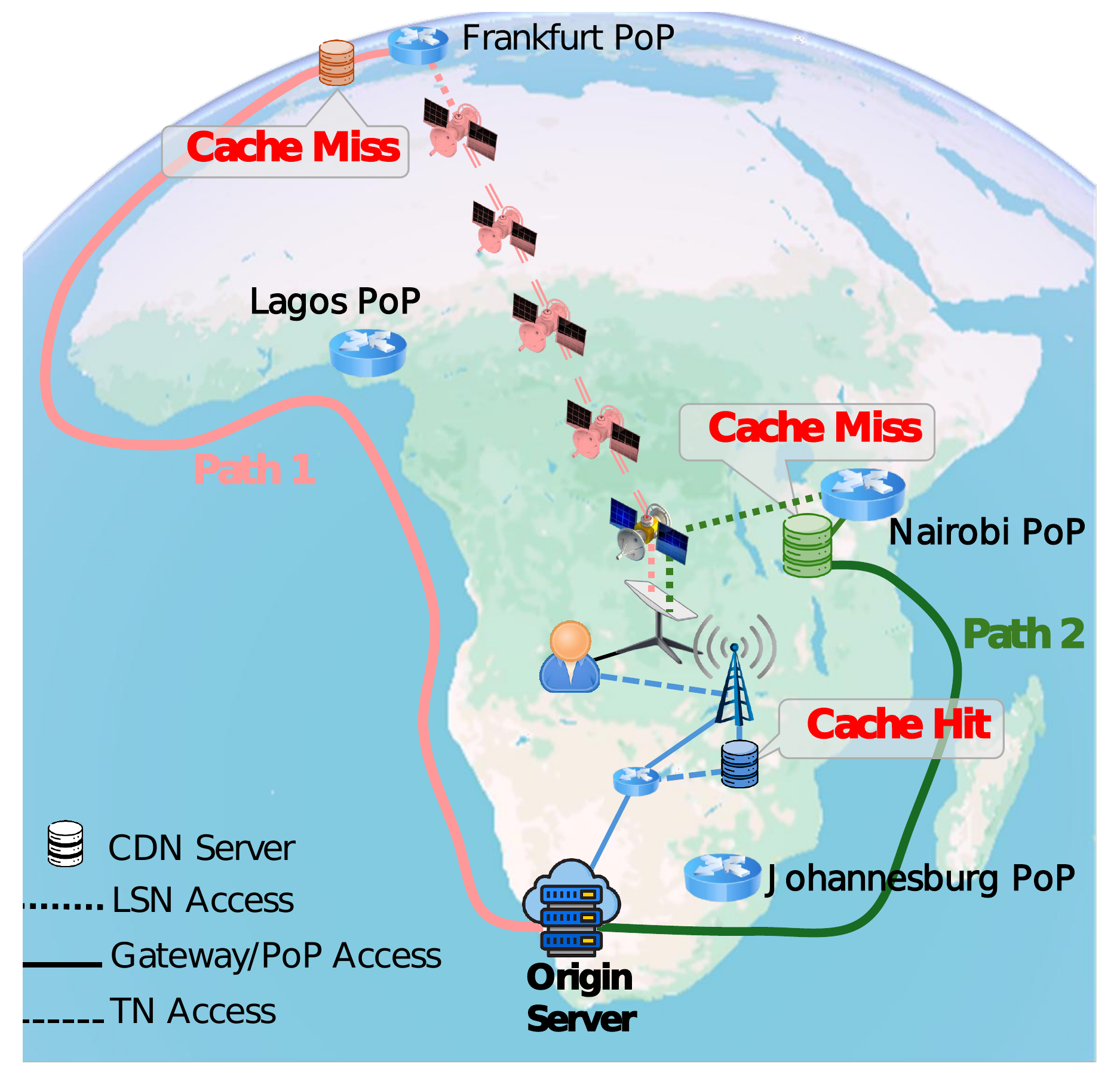}
    \caption{TN subscribers reach local CDN edges hosting locally relevant content, while LSNs must first traverse to distant PoPs before accessing CDN infrastructure. CDN mapping techniques select caches near the PoP resulting in cache misses and increased latency to fetch content from origin.}
    \label{fig:starlink-bentpipe}
    \Description{Overview TN vs LSN}
    \vspace*{-1.5em}
\end{figure}

While existing research extensively characterizes network-layer performance of LSNs~\cite{michel2022,kassem2022,pan2023measuring,dissectingleo,tiwari2023t3p,garcia2023multi,multifacetedStarlink-www24,izhikevich2023democratizing,laniewski2025measuring,ullah2025impact}, these studies overlook how LSN architecture fundamentally disrupts content delivery assumptions that have underpinned CDN operations for decades.
Unlike terrestrial networks where users typically connect to nearby CDN cache servers hosted by local ISPs, LSN users' traffic emerges at distant points-of-presence (PoPs) that may be thousands of kilometers away~\cite{wang2024large,multifacetedStarlink-www24}.
\Cref{fig:starlink-bentpipe} illustrates this architectural mismatch through a concrete example from our measurements in African sub-continent.
Before January 2025, Starlink users in Zimbabwe (ZW) were routed via inter-satellite links (ISLs) to Frankfurt PoP in Germany (DE), approximately 7000 km away (\colorbox{pink}{Path 1}).
This remote PoP assignment had cascading effects: DNS resolvers were selected based on the Frankfurt location, CDN mapping logic placed these users at European caches, and cache misses due to the lack of African content cached in EU servers resulted in intercontinental path traversals to origin server in Africa despite users being physically close to the server.
Even after Starlink's PoP reassignment to Kenya (KE) in Jan'25 (\colorbox{green}{Path 2}), while reducing the user-to-PoP distance to approximately 1000 km, performance remained suboptimal compared to TNs that accessed local CDN caches within Zimbabwe itself (unlike KE servers which observed large number of cache misses for ZW content).
This underscores how PoP assignments trigger a cascade of suboptimal decisions across multiple layers of content delivery infrastructure.
Yet no prior work has systematically measured these compound effects or their impact on user-perceived performance.

In this paper, we present the \emph{first} comprehensive, global measurement study of end-to-end content delivery performance over Starlink, systematically decoupling the impacts of PoP assignment, DNS resolution, and CDN request-mapping on user-perceived web performance.
While prior work has characterized Starlink's network performance~\cite{liz2024netflix} and identified conceptual CDN-LSN mismatches~\cite{leoCDN-hotnets2024}, no study has decomposed how these architectural differences compound across multiple layers affect actual content delivery.
Through multi-layer measurements spanning 145 countries and encapsulating Starlink's infrastructure expansion in 2025, we quantify how satellite-based Internet fundamentally disrupts the underlying assumptions of modern content delivery networks.
Specifically, we make the following contributions:

\noindent
\textbf{(1)} We conduct a longitudinal \emph{two years} (Nov 2023 - Sept 2025) \textbf{global measurement} campaign combining \emph{passive analysis} of Cloudflare AIM~\cite{cloudflare-aim} and M-Lab~\cite{mlab} datasets with \emph{active measurements} from 99 Starlink-connected RIPE Atlas probes (6.1~M traceroutes, and 10.8~M DNS queries).
Our multi-layer analysis reveals three distinct Starlink performance regimes: content-rich PoPs where the latency of satellite links dominates (80--90\% of RTT), sparse-edge regions with inflated PoP-to-CDN segments, and remote-PoP scenarios exceeding 200~ms total latency (see \cref{subsec:content-delivery-networks}).
We find that anycast CDNs (Cloudflare) consistently outperform DNS-mapped alternatives (Akamai and CloudFront) by $\approx$18~ms on average, with gaps exceeding 100~ms when resolver mislocalization occurs (such as Manila PoP users reaching Hong Kong DNS resolvers).
DNS cache hit rates vary significantly across providers, with Cloudflare showing the highest rates, while resolution times in Africa and South America suffer from both elevated resolver latency and increased cache misses (see \cref{subsec:domain-name-system}).
Compared to TNs, Starlink exhibits 2--3$\times$ higher median latencies in well-provisioned regions but surprisingly outperforms TNs in selected African countries where satellite paths bypass congested terrestrial routes to Europe.
This comprehensive characterization demonstrates that PoP distance explains up to 50\% of performance variance, with DNS and CDN layer misalignments accounting for the remainder.

\noindent
\textbf{(2)} We perform \textbf{controlled measurement} across three continents, operating Starlink dishes in Ghana, Zambia, Canada (3 sites), and Germany to conduct systematic DNS and HTTP GET measurements every two hours for six months.
This infrastructure uniquely captures multiple Starlink PoP reassignments in early 2025 as natural experiments, revealing stark regional differences in impact.
In Africa, where terrestrial infrastructure remains limited, relocating users from European to local PoPs (Nairobi, Johannesburg) yields dramatic improvements with median page-fetch times dropping 60\% and cache hit rates increasing from 60\% to 85\% (see~\cref{sec:africa_case}).
However, similar PoP reassignments in Canada show minimal effects, as dense CDN deployments already exist near both old and new PoPs.
Our measurements further showcase that reassignments to CDN-sparse regions increase server location variance by up to 3$\times$, with users now distributed across diverse edge locations rather than consistently hitting content-rich hubs (see~\cref{sec:other_reassignments}).
This differential impact demonstrates that infrastructure proximity benefits depend fundamentally on regional CDN maturity.

Our findings reveal \ul{fundamental architectural misalignments between satellite-based Internet access and terrestrial content delivery infrastructure}, with implications for multiple stakeholders.
For \emph{CDN operators}, our results demonstrate the urgent need for satellite-aware request-mapping techniques that account for the geographic disconnect between user location and PoP egress, particularly as LSNs expand in under-served regions where traditional geolocation assumptions fail most severely.
For \emph{LSN operators}, we provide empirical evidence that strategic PoP placement in content-rich locations yields greater performance improvements than simply minimizing geographic distance to users.
For \emph{researchers}, our study provides a foundation for exploring novel architectures such as in-orbit caching~\cite{leoCDN-hotnets2024,starcdn-sigcomm2025}, satellite-native CDN protocols, and request-mapping techniques that leverage ISLs topology awareness to optimize content delivery paths~\cite{spache2025}.
For \emph{policymakers and network planners} in developing regions, our measurements quantify the hidden costs of poor peering and limited local content hosting, supporting broader initiatives to ``keep local content local''~\cite{internetsociety_50-50_vision_2025} through targeted infrastructure investment.
To foster reproducibility and enable future research on this rapidly evolving ecosystem, we will release our measurement datasets, analysis code, and controlled measurement methodology publicly upon acceptance.

%% file: sections/methodology.tex
\section{Measurement Methodology}\label{sec:methodology}

\subsection{Passive Measurements} \label{sec:methodology:passive}

\noindent\textbf{Cloudflare AIM.}
We analyze the Cloudflare Aggregated Internet Measurements (AIM) dataset~\cite{cloudflare-aim}, which captures client-initiated speed tests to Cloudflare's global CDN infrastructure.
The AIM test suite employs a browser-based approach that initially establishes baseline network conditions by measuring idle latency with small TCP probes.
It then conducts progressive download and upload tests over HTTPS to measure throughput and latency under load~\cite{cloudflare_network_2025}.
These tests reflect real-world CDN performance as experienced by actual users, as Cloudflare's anycast mapping naturally routes each client to what the infrastructure determines the ``optimal'' edge server based on BGP paths and network conditions.
We extract Starlink measurements by filtering for ASN 14593 and ASN 45700, yielding 255K speed tests from November 2023 to September 2025 across 145 countries where Starlink operates.
Each record contains city-level geolocation for both the client and the Cloudflare edge server, enabling analysis of how PoP assignments influence Cloudflare CDN server selection over time~\cite{CloudflareSpeedTest}.
Since anycast routing and Cloudflare's internal load balancing can direct the same client to different servers across measurements, we use the median idle latency per location to identify the typical CDN server assignment for that network position.
To establish a terrestrial baseline and comparative analysis, we extract measurements from the three highest-volume ISPs in each country (see \Cref{subsec:content-delivery-networks} and \Cref{sec:africa_case}).

\smallskip
\noindent\textbf{{Measurement Lab (M-Lab).}}
While Cloudflare AIM measurements provide a global overview of CDN performance, they lack path-level visibility into the Starlink connection, which is crucial for understanding the impact of Starlink's topological routing on content delivery.
For this, we turn to M-Lab speed tests~\cite{mlab-pub,mlab-platform}. 
M-Lab is Google-supported open measurement platform that conducts browser-based measurements between clients and over 500 servers distributed across 60+ metropolitan areas worldwide.
M-Lab tests perform multi-stream TCP throughput measurements while simultaneously collecting packet-level diagnostics and conducting reverse traceroutes from the measurement server back to the client~\cite{mlab-platform}.
We analyze these reverse traceroutes from M-Lab servers to Starlink clients during the same period as our AIM analysis (Nov 2023 - Sept 2025), focusing on identifying the last public Internet hop before entering Starlink's network to determine the active PoP for each user.
By tracking these PoP assignments across time and geography, we can correlate Starlink's evolving infrastructure topology with the CDN performance variations observed in AIM data.
Together, AIM's CDN metrics and M-Lab's path topology enable us to decompose the global end-to-end latency of CDN performance over Starlink into its constituent components: the satellite segment (user-to-PoP) and the terrestrial segment (PoP-to-CDN).

\subsection{Active Measurements} \label{sec:methodology:active}

\textbf{RIPE Atlas.}
RIPE Atlas is a measurement platform that the networking research community commonly employs for conducting measurements~\cite{ripe-atlas}.
We utilized 99 Starlink-connected RIPE Atlas probes (ASN 14593) across 32 countries.
We orchestrated fine-grained active measurements from these probes to complement our passive analysis, which provided us limited control over measurement targets and timing.
Our measurement campaign spanning a total of fifteen weeks between March to September 2025 included \emph{DNS} and \emph{traceroute} measurements to carefully selected CDN-hosted websites from the Tranco top-2~K list~\cite{Tranco}.
We curated targets to equally represent the three major CDN providers: Cloudflare~\cite{cloudflare_network_2025}, Akamai~\cite{akamai_global_infrastructure_2025}, and CloudFront~\cite{aws_global_infrastructure_2025} (see \Cref{table:website-list} for the complete list).

For DNS measurements, we performed two query types from each probe:
(1) CHAOS class queries to Cloudflare, Google, and Quad9 resolvers for the \texttt{id.server} TXT record, which returns resolver location identifiers (e.g., \url{res760.}\texttt{\underline{sea}}\url{.rrdns.pch.net});
(2) A-record queries for five websites per CDN provider with recursion disabled (RD=0), forcing resolvers to return only cached responses, thereby allowing us to measure cache hit rates and response times without recursive resolution overhead.
For path measurements, we conducted TCP traceroutes (with UDP fallback) to all target websites, resolving domains locally before each measurement.
We geolocated intermediate hops using reverse DNS semantics embedded in PTR records (e.g., \url{ae0.}\texttt{\underline{lax}}\url{.netarch.akamai.com})~\cite{conext2021learning}, validated against IPInfo's geolocation database~\cite{iptocountryipinfo}.
Hops within 2~ms of identified geographic markers were tagged with the corresponding location, enabling precise PoP and CDN server identification.
All measurements were repeated every 15 minutes, producing 6.1~M traceroutes and 10.8~M DNS queries.
We conducted identical measurements from terrestrial RIPE Atlas probes from the same regions to establish baseline comparisons.

\smallskip
\noindent\textbf{Controlled Vantage Points.}
To complement our global probe measurements with application-level performance metrics, we deployed six controlled Starlink user terminals across three continents: 3$\times$ Canada, 1$\times$ Germany, 1$\times$ Ghana, and 1$\times$ Zambia.
These controlled vantage points (VPs) enable us to conduct identical measurements across diverse geographic and infrastructure conditions while capturing real web browsing performance through HTTP fetches.
From March to September 2025, each VP executed following measurements every two hours: (i) replicating DNS and traceroute tests to the same website endpoints as our RIPE Atlas probes, and (ii) fetching complete webpages from target list (\cref{table:website-list}) using \texttt{curl} to capture time-to-first-byte (TTFB) and extract cache status directly from HTTP response headers.
Overall, we collected 43~K DNS CHAOS queries, 819~K DNS A-record queries, 512~K traceroutes, and 523~K HTTP GET fetches.
For Cloudflare-hosted websites, we extracted cache hit status from the \texttt{CF-Cache-Status} header and CDN server location from the \texttt{CF-Ray} header, which encodes the three-letter IATA airport code of the serving edge.
CloudFront similarly provides cache status via \texttt{X-Cache} and server location through the \texttt{X-Amz-Cf-Pop} header.
Akamai, however, does not expose cache information in standard response headers.
The request header \texttt{Pragma: akamai-x-cache-on} has to be set in order to receive cache status in the \texttt{X-Cache} response header, while CDN server locations are not explicitly provided.
Note that our experiments capture multiple Starlink PoP reassignments during the period from these locations: Zambia terminal transitioned from Nairobi to Johannesburg in April 2025, while our Canada deployments cover mappings to both in-country (Calgary and Montréal) and US (Seattle and New York City) PoPs.
Germany and Ghana terminals remained stable at Frankfurt and Lagos PoPs, respectively.

%% file: sections/global.tex
\section{Global Content Delivery Overview}\label{sec:global_overview}

\begin{figure}[t]
  \includegraphics[width=0.8\columnwidth]{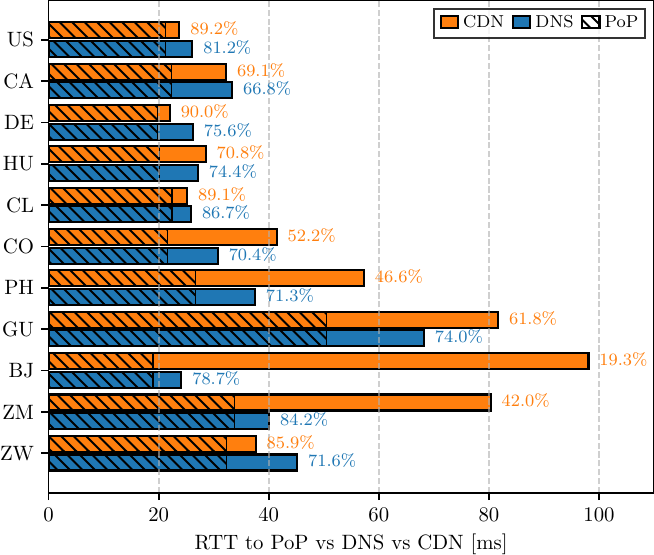}
  \vspace{-1em}
  \caption{CDN latency breakdown over Starlink from selected countries: median minRTT to CDN caches (in \textcolor{orange}{orange}) and DNS resolvers (in \textcolor{blue}{blue}). Hatched sections isolate the PoP segment, with percentages marking end-to-end RTT share.}
  \label{fig:global_rtt_to_pop_vs_cdn}
  \vspace{-1.5em}
\end{figure}

A typical webpage fetch begins with DNS resolution of the domain name; once resolved, the browser connects to the CDN edge and retrieves the page.
Consequently, both the DNS resolution time (which includes latency to the DNS resolver) and the latency to the CDN edge are key contributors to user-perceived performance~\cite{doh-webperf,goenka2022towardscdn,zolfaghari2020cdnsurvey,fan2015assessing}.
For Starlink users, each of these components can be further decomposed into two segments: satellite traversal from user to the assigned Starlink PoP, and the terrestrial path from the PoP to the (resolver or CDN edge) endpoint (see~\Cref{fig:starlink-bentpipe}).
Using traceroutes from RIPE Atlas and our controlled VPs, we isolate each component in end-to-end content fetch, i.e. latency to Starlink PoP (last public Internet hop), DNS resolver, and CDN edge.
We take the \emph{minimum} RTT (minRTT) recorded across our targeted websites (\cref{table:website-list}) and then compute the median across vantage points.
\Cref{fig:global_rtt_to_pop_vs_cdn} shows the breakdown of web fetch latencies over Starlink across selected countries.
Blue bars denote the latency to the DNS resolver, orange bars shows the latency to the CDN edge,  and the hatched overlay on both bars shows the median latency from client to the mapped PoP.
We aggregate all CDN providers (Cloudflare, Akamai, and CloudFront) and DNS providers (Cloudflare, Google, and Quad9) measurements here, and discuss operator-specific performance in~\Cref{subsec:content-delivery-networks}.

\smallskip
\noindent \textbf{Content-rich PoPs.} In countries such as the United States (US), Germany (DE), and Chile (CL), the serving PoP sits within the same country. Moreover, these regions also have a dense CDN and DNS infrastructure.
As a result, we observe that the satellite path (user-to-PoP segment) dominates the end-to-end latency budget ($\approx$80--90\,\%), and the latency from PoP to CDN/DNS server is minimal.
Starlink PoPs are often co-located with regional IXPs~\cite{wang2024large,multifacetedStarlink-www24}, allowing direct peering with major ISPs and CDN providers.
In NA and EU, the ``bent-pipe'' latency baseline is $\approx$25--30\,ms.
With nearby edges and resolvers, this baseline largely determines user-visible delay.
Comparatively, terrestrial infrastructure in CL is less developed, the Santiago PoP is well provisioned with CDN edges\footnote{\url{https://www.peeringdb.com/ix/1514}}, so Starlink users there achieve CDN latencies comparable to NA and EU.

\smallskip
\noindent \textbf{Sparse-edge regions.} In several countries like Colombia (CO), Hungary (HU), Benin (BJ), and the Philippines (PH), users enjoy low latency to a nearby PoP yet still face elevated PoP-to-CDN/DNS latency because their requests are mapped to caches in neighboring metros or countries.
For example, Starlink users in BJ are associated with the Lagos (LOS) PoP, but often connect to CDN edges in Johannesburg (JNB) or occasionally in EU.
As a result, users experience significantly higher latency-to-content ($\approx$100\,ms), with only 19\% attributed to the latency-to-PoP segment.
Interestingly, BJ users connect to DNS resolver instances predominantly in Lagos, yielding DNS latencies nearly identical to their PoP latency.
Yet this proximity advantage is lost when the resolver maps these requests to distant CDN server due to limited regional CDN availability.

\smallskip
\noindent \textbf{Remote-PoP scenarios.}
In regions like Zambia (ZM), Zimbabwe (ZW), and Madagascar (MG) (not shown), Starlink users connect to remote PoPs in other parts of Africa due to absent local ground infrastructure.
These users experience significant user-to-PoP latency overhead before exiting to the public Internet.
Once at the PoP, both DNS resolver and CDN latencies can be relatively short, but the aggregate already exceeds 80\,ms due to Starlink's internal traversal.
Note that prior to Jan 2025, users from these countries were often assigned to EU, resulting in even higher latencies due to longer satellite paths and increased cache misses.
We discuss the impact of these infrastructural enhancements in in~\Cref{sec:africa_case}.

\subsection{CDN Performance} \label{subsec:content-delivery-networks}

\begin{figure}[!t]
  \includegraphics[width=\columnwidth]{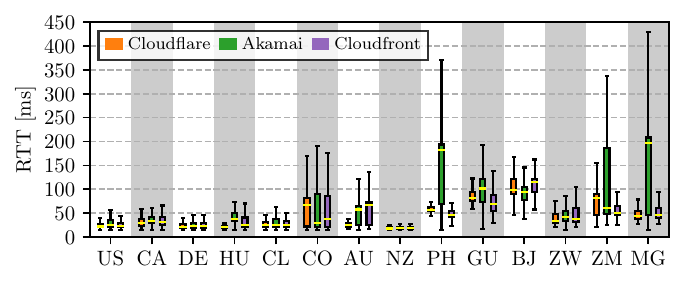}
  \vspace*{-3em}
  \caption{Access latencies from Starlink subscribers to CDN servers of Cloudflare, Akamai and CloudFront.}
  \label{fig:rtt_to_three_cdn_providers}
  \Description{Access latencies to CDN servers of Cloudflare, Akamai and CloudFront from Starlink subscribers in selected countries.}
  \vspace*{-1em}
\end{figure}

We benchmark CDN performance for Starlink across Cloudflare, Akamai, and CloudFront using traceroutes from RIPE Atlas probes and our controlled nodes (see \Cref{fig:rtt_to_three_cdn_providers}).
For each path, we geolocate both the Starlink PoP and the CDN hop as described in~\Cref{sec:methodology:active}.
Since Cloudflare employs anycast~\cite{calder2015analyzing} while Akamai and CloudFront use DNS-based request redirection~\cite{endusermapping-sigcomm15}, our analysis provides a uniform baseline for comparing across these different mapping schemes.

\begin{figure}[t]
  \includegraphics[width=\columnwidth]{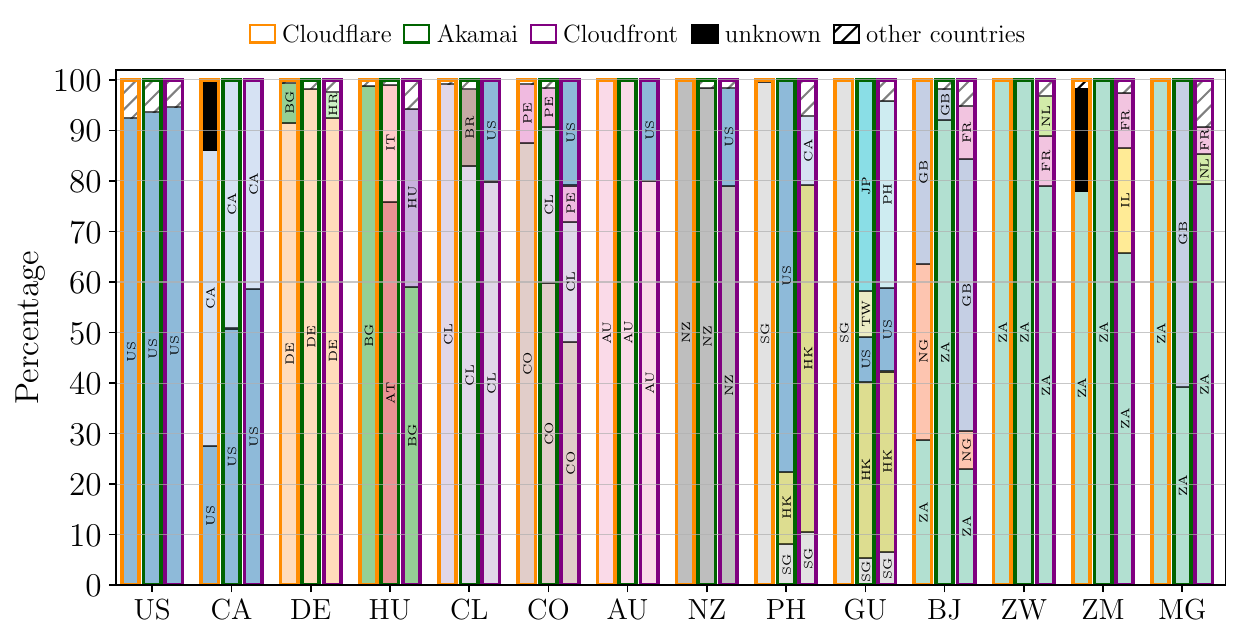}
  \vspace*{-2.5em}
  \caption{Provider-specific CDN mapping across Starlink. Each bar shows the distribution of CDN server locations selected by the provider's mapping (labels inside bars) across Cloudflare (left), Akamai (center) and Cloudfront (right).}
  \label{fig:cdn_locations_three_providers}
  \vspace*{-2em}
\end{figure}

In regions with \emph{dense CDN infrastructure deployment}, e.g., NA and EU, all CDN providers showcase similar latencies because edges are commonly co-located with, or near, the serving Starlink PoP.
Elsewhere, the provider differences emerge.
Cloudflare consistently achieves the lowest latencies $\approx$18\,ms lower than Akamai and $\approx$6\,ms lower than CloudFront on average, due to its anycast mapping that naturally steers traffic to the nearest available edge.
In contrast, DNS-based providers (Akamai and CloudFront) exhibit higher variance, with Akamai particularly vulnerable to resolver mis-localization despite its comparable global footprint.
The difference across providers is most visible where the PoP's surroundings are \emph{content-sparse}, e.g., users in Colombia (CO), Benin (BJ), the Philippines (PH), and Guam (GU) often traverse to edges in a different metro/country, inflating end-to-end latency.
For instance, PH users experience RTTs of $\approx$57\,ms to Cloudflare and $\approx$46\,ms to CloudFront servers, but $\approx$180\,ms to Akamai servers.
To understand this further, we analyze the CDN mapping patterns across providers in~\Cref{fig:cdn_locations_three_providers}, which indicates distribution of CDN edge server locations selected by each provider.
We confirm that for content-rich regions (US, DE), all providers predominantly select local edges.
For content-sparse regions, however, server selection varies widely and is dependent on provider-specific infrastructure and mapping logic.
For instance, Akamai maps PH users to US servers 75\% of the time, while Cloudflare and CloudFront predominantly select Singapore (SG) and Hong Kong (HK) edges, much closer to Manila (MNL), explaining the elevated latency in~\Cref{fig:rtt_to_three_cdn_providers}.
GU users exhibit similar patterns: after the Tokyo-to-Manila PoP reassignment, RTTs to Cloudflare and CloudFront drop to $\approx$70\,ms, yet Akamai remains elevated because it continues mapping requests to JP rather than edges near Manila (details in \cref{sec:other_reassignments}).

For South Eastern African countries, Zimbawe (ZW) users experience slightly lower latencies than Zambia (ZM) or Madagascar (MG).
ZW users exiting via the Johannesburg (JNB) PoP benefit from short PoP--CDN paths across all providers, as~\Cref{fig:cdn_locations_three_providers} shows near-universal selection of South Africa (ZA) edges.
MG users also connect to the JNB PoP, but while Cloudflare anycast maps the user requests to nearest ZA servers, DNS-based server mapping scheme followed by Akamai and CloudFront often select EU edges (e.g., GB, NL, FR) due to resolver mis-localization, leading to higher latencies.
We discuss DNS resolution using public resolvers, and its impact on server selection further in~\Cref{subsec:domain-name-system}.
Our ZM measurements also reveal infrastructure-dependent variations: the RIPE Atlas probe connects via Nairobi (NBO) while our controlled dish uses JNB.
The NBO-connected ZM node suffers higher CDN latencies due to both greater distances to CDN servers and the region's sparser infrastructure compared to JNB.
The subsequent NBO-to-JNB reassignment improved latencies for affected users (see \Cref{sec:africa_case}).
BJ users, connected to the Lagos (LOS) PoP, consistently face $>$90\,ms CDN latencies due to sparse infrastructure surrounding LOS which forces connections primarily to ZA and NG, or even occasional EU edges (GB, and FR).

We also find that terrestrial ISPs achieve consistent performance with median latencies of 19-20\,ms and tight distributions, while Starlink exhibits higher medians ($\approx$50\,ms) and greater variance. (plot not shown due to space restrictions).
This disparity stems from (i) the mandatory satellite traversal, and (ii) the cascading effects of remote PoP assignments on CDN selection.
In content-rich regions (US, CA, DE, HU, and CL), where CDN edges co-locate with Starlink PoPs, the satellite segment accounts for 25--30\,ms latency overhead with minimal variance.
On the other hand, terrestrial users connect directly to local CDN caches, while Starlink users must first traverse to the PoP before reaching the same infrastructure (as depicted in~\Cref{fig:starlink-bentpipe}).
In sparse-edge regions (e.g., PH, BJ), Starlink's performance becomes more variable.
While users reach nearby PoPs quickly, subsequent mapping to CDN servers in neighboring countries inflates both median latency and variance.
PH users, for instance, observe a long latency tail due to being mapped to edges in HK, US and CA.
Notably, while~\cite{multifacetedStarlink-www24} documented baseline latency improvements for PH users after the Manila PoP deployment in late 2023, \emph{our measurements reveal that these network-layer gains fail to translate to improved web performance due to persistent misalignment between CDN infrastructure and Starlink network}.
Surprisingly, we also find that this architectural mismatch occasionally benefits Starlink users.
In BJ and MG, terrestrial ISPs often hairpin traffic to European caches over 7000\,km away, whereas Starlink's Lagos (LOS) PoP enables intra-continental routing to South African edges, reducing end-to-end latency by 30--60\,ms compared to terrestrial paths.
We explore this counter-intuitive advantage further in~\Cref{sec:africa_case}.

\begin{tcolorbox}[title=\textit{Takeaway \#1}, enhanced, breakable]
  CDN architecture fundamentally determines Starlink performance.
  Anycast CDNs (Cloudflare) maintain consistent low latency by naturally selecting PoP-adjacent edges, while DNS-based CDNs (Akamai and CloudFront) suffer from resolver mis-localization, with latency penalties exceeding 100\,ms in sparse-edge regions.
  Starlink can paradoxically outperform terrestrial networks where inefficient peering creates longer paths, but only when CDN providers accurately track Starlink's evolving GeoFeed and PoP topology~\cite{starlink-geoip-csv}. Without PoP-aware mapping, infrastructure improvements fail to translate to application-level gains.
  \end{tcolorbox}

\subsection{DNS Performance} \label{subsec:domain-name-system}

DNS holds a critical yet sometimes underestimated role in web performance in two aspects: (i) resolution time contributes directly to page startup delay, and (ii) for DNS-mapped CDNs, the resolver's viewpoint steers which CDN edge is selected.
While we previously highlighted misalignment between Starlink PoPs and CDN servers providing requested content, DNS resolution time can further exacerbate this effect when resolvers are not co-located with the user's PoP.
We investigate this by focusing on three popular public DNS resolvers: Cloudflare (\texttt{1.1.1.1}), Google (\texttt{8.8.8.8}), and Quad9 (\texttt{9.9.9.9}), with the former two used by Starlink by default.
All three resolvers use IP anycast to direct DNS resolution requests to the nearest available resolver instance.
Note that our target domain list remains the same as in \cref{subsec:content-delivery-networks} (\Cref{table:website-list}).
To learn DNS resolver latency and location, we follow \texttt{RFC4892}~\cite{RFC4892} by sending \texttt{CHAOS} class DNS queries for the \texttt{id.server} \texttt{TXT} record (see \cref{sec:methodology}).

\begin{figure}[!t]
  \includegraphics[width=\columnwidth]{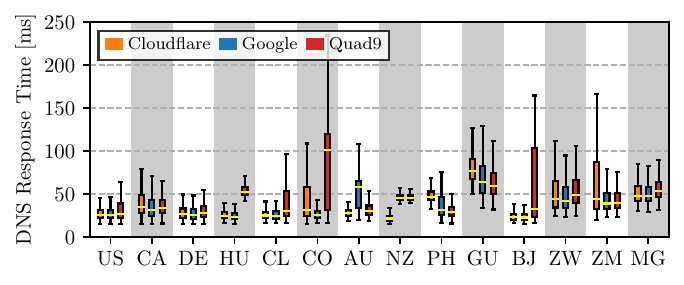}
  \vspace*{-2.5em}
  \caption{DNS response times across Cloudflare, Google and Quad9 public resolvers for Starlink users.}
  \label{fig:dns_response_time_three_providers}
  \Description{DNS response times across Cloudflare, Google and Quad9 public resolvers for Starlink users.}
  \vspace*{-2em}
\end{figure}

\Cref{fig:dns_response_time_three_providers} shows the latencies to the DNS resolvers for all providers.
Note that resolver response time signifies the baseline cost of \emph{reaching} DNS infrastructure and remains consistent regardless of the domain being queried, providing a clean measure of resolver-PoP alignment.
Globally, Cloudflare and Google achieve the lowest median latency at $\approx$28\,ms, while Quad9 lags at $\approx$33\,ms.
Latencies are lowest when a resolver PoP is co-located with the Starlink home PoP, as in the case of US, DE, and CL.
However, for PoPs located in regions where terrestrial infrastructure is not yet mature, e.g., Manila (PH), Lagos (NG), Nairobi (KE), Bogota (CO) and Sofia (BG), Starlink users are resolved by servers significantly further away from PoP location leading to higher DNS response times.
For instance, PH and GU users assigned to the MNL PoP, connect to Cloudflare resolver instance in HK ($\approx$ 1100 kms away), adding 18--20~ms median latencies compared to TN users.
We also find provider-specific differences as Google and Quad9 more frequently select local MNL resolvers for PH users.
In HU, traffic exits via Sofia PoP, but Quad9 often routes to DE (Frankfurt), doubling DNS delay ($\approx$50~ms) compared to Cloudflare and Google.
Similarly, in CO, users achieve $\approx$25\,ms to Google resolvers in Bogota where the PoP resides.
For the same users, Cloudflare latencies increase to $\approx$36\,ms when connecting to Lima (25\% of queries), while Quad9 adds $\approx$82\,ms when routing to Santiago (62\% of queries).
While it may appear that these stark differences majorly stem from provider infrastructure density across the globe as Cloudflare and Google typically outperform Quad9.
However, \cite{dnsoh-imc21} showed Cloudflare and Quad9 DNS infrastructure to be significantly more widespread than Google's, which has a smaller footprint and is more concentrated in NA and EU.
Upon further investigation, we find that, in addition to infrastructure distribution, DNS resolvers can mis-localize Starlink users due to inaccurate Starlink GeoIP feeds, which can misrepresent a user's PoP location~\cite{baraya_starlink_infrastructure_report_2025}.
For instance, AU probes assigned with the Perth PoP connect to Google resolver instances in Melbourne and Sydney ($>$2700 kms away), while queries to Cloudflare \& Quad9 are resolved in Perth locally.

We now turn to end-to-end DNS resolution time, which includes both resolver RTT and the query processing time.
Note that the latter is affected by whether the query is cached at the resolver or requires a recursive lookup, and is therefore dependent on domain popularity and temporal locality.
As discussed in \cref{sec:methodology}, we track cache hits at resolver by issuing A-record queries with \texttt{RD=0}: a reply indicates a cache hit; no data/\texttt{REFUSED} indicates a miss (no recursion).
\Cref{fig:dns_resolution_times_three_providers_appendix} in ~\Cref{app:dns-resolution-latency} shows the results.
Three key patterns emerge.
\emph{First}, cache hit rates vary substantially across providers and regions.
Cloudflare maintains the highest global hit rate at 78\%, compared to Google's 65\% and Quad9's 58\%.
However, these rates degrade for region-specific content: African domains show 20--25\% lower hit rates across all providers, with Quad9 dropping below 40\% for local African content.
This disparity reflects both cache size differences and query distribution biases toward NA and EU content~\cite{randall2020trufflehunter}.
\emph{Second}, resolution times scale non-linearly with resolver distance when misses occur.
We find that while US and DE users maintain sub-50\,ms resolution times even with misses, African users (BJ, ZM, and MG) face 200--300\,ms penalties.
Each recursive hop multiplies the base latency, with a miss requiring authoritative lookup plus CNAME resolution, which adds 3--4 round trips.
For ZM users with $\approx$80\,ms resolver latency, this translates to >200\,ms additional overhead per miss.
\emph{Third}, resolver misalignment creates cascading impacts for DNS-based CDNs.
When CO users query through Santiago-based Quad9 resolvers (82\,ms away), not only does resolution take longer, but Akamai and CloudFront return edges near Santiago rather than Bogota.
This compounds to 150\,ms DNS resolution plus 100\,ms to the wrong CDN edge, i.e., $\approx$250\,ms, before content transfer begins.
Similarly, PH users reaching HK resolvers face both elevated resolution times and subsequent mapping to HK/SG edges rather than MNL-local caches.

\smallskip
\noindent
\textbf{Resolver-driven CDN mapping (for DNS-mapped CDNs).}
Akamai and CloudFront select CDN caches via DNS, with server selection determined by the resolver's location rather than the user's actual position.
When the resolver is not co-located with the Starlink PoP, these CDNs return unicast IPs near the resolver rather than near the PoP~\cite{perdices2022satellite}, adding unnecessary terrestrial segments after the satellite hop.
Our measurements reveal this effect clearly: PH users attached to the MNL PoP connect to DNS resolvers in HK and Taiwan (TW), $\approx$1100--1500\,km away.
When these distant resolvers query Akamai or CloudFront, the CDNs return edge addresses near HK or TW, forcing Starlink traffic to traverse across the South China Sea.
This resolver-induced detour adds 40--60\,ms beyond what MNL-local edges would require.
We also observe ``\emph{resolver bias}'' where different public resolvers in the same location return different CDN mappings.
When querying from PH probe through Google's MNL resolver, 10--20\% of Akamai responses direct users to NZ edges (> 7000\,km away).
The same queries through Quad9's MNL instance consistently return Southeast Asian edges.
This inconsistency suggests that CDN mapping algorithms weigh resolver identity differently, potentially using historical traffic patterns or peering relationships specific to each resolver network.
These biases compound latency unpredictably: the same Starlink user accessing the same website may experience 50--150\,ms variation depending solely on which public resolver handles their query.

\begin{tcolorbox}[title=\textit{Takeaway \#2}, enhanced, breakable]
  Resolver-PoP misalignment triggers cascading impacts: 20--80\,ms direct penalties compound through 3--4$\times$ amplification during cache misses, pushing resolution times to 250--300\,ms in Africa.
  DNS-based CDNs inherit these mis-alignments, selecting edges near resolvers rather than users.
  With regional cache hit rates 20--25\% lower and resolver bias adding 50--150\,ms variation, DNS latency often exceeds content transfer time itself in under-served regions.
  \end{tcolorbox}

%% file: sections/africa.tex
\section{Case Studies}

\subsection{African Infrastructure Expansion} \label{sec:africa_case}

\begin{figure}[!t]
    \includegraphics[width=0.85\columnwidth]{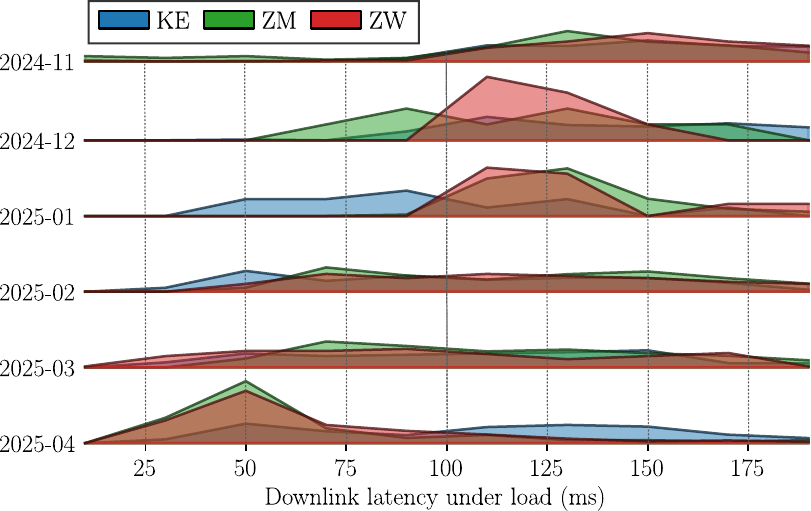}
    \vspace*{-1em}
    \caption{Starlink latency distributions to Cloudflare from selected African countries before (EU PoP) and after (local AF PoP) remapping in January 2025.}
    \label{fig:af_countries_latencies}
    \Description{Starlink latency distributions from selected African countries to best Cloudflare server location. In January 2025, the AF countries were remapped to PoPs within AF instead of EU.}
    \vspace*{-2.5em}
\end{figure}

We draw specific focus on Africa, where sparse terrestrial fiber and heavy reliance on a few submarine routes to EU creates high latency and fragility~\cite{carisimoHopAwayEverywhere2023}.
In such settings, LSNs have a natural advantage over TNs by bypassing substandard terrestrial segments and exiting closer to content~\cite{leoCDN-hotnets2024,starcdn-sigcomm2025}.
However, recent studies on Starlink operations have shown that, apart from the Lagos (LOS) PoP in Nigeria, which was the only African PoP serving nearby countries before 2025, most African Starlink users suffered from high latencies.
This issue arose from the reliance on ISLs to connect to PoPs located in EU, notably in Frankfurt (FRA), DE~\cite{multifacetedStarlink-www24}.
This mapping inflates both the user-to-PoP leg and the PoP-to-CDN leg (toward EU caches), degrading CDN-dependent application performance.
For example, Netflix streaming sessions saw 2$\times$ as much rebuffering compared to users in NA and EU~\cite{liz2024netflix}, and web browsing of popular websites incurred $\approx$200~ms higher latencies~\cite{leoCDN-hotnets2024}.

Starlink has been actively investing in expanding its ground infrastructure in AF, adding two new PoPs in NBO and JNB in early 2025, along with several new ground stations in the region~\cite{starlink-gs-map}.
From M-Lab measurements, we confirm that Starlink performed AF-wide PoP reassignments in January and April 2025 after which the majority of AF users are associated with local PoPs in LOS, NBO or JNB.
While these deployments should markedly reduce user-to-PoP latency across Africa, their effect on latency-to-content remains unclear, creating a natural before/after experiment to quantify the end-to-end impact.
\Cref{fig:af_countries_latencies} illustrates the distribution of downlink latencies-under-load from selected AF countries to Cloudflare CDN.
Note that Cloudflare hosts servers in 25 of 54 African countries~\cite{cloudflare_network_2025}, and our analysis uses the AIM dataset to track latencies before and after the January and April 2025 PoP reassignments.
Following these changes, the latency distributions for most African countries have improved by $\approx$2$\times$, with latencies in May 2025 comparable to those reported in the NA, and EU (see~\Cref{sec:global_overview}).
\Cref{fig:cdf_af_cloudflare_all_countries} (in~\Cref{app:africa-cloudflare-aim}) showcases latency-to-content before and after PoP reassignment from more African countries.
The impact is most pronounced for South Eastern African countries, such as ZM, ZW and MG.
Latencies from these regions were initially affected by highly inconsistent long-haul ISL connections to the FRA PoP before being routed to CDN servers across multiple EU countries, such as Germany (DE), Spain (ES), Switzerland (CH), and Portugal (PT).
Following the PoP reassignments, the majority of Starlink users in South Eastern African countries are mapped to local CDN servers within the same country as the PoP, predominantly around JNB.
This highlights both the benefit of localization and a content-concentration effect as many caches and popular objects are still clustered near JNB.
By contrast, NG shows little change: latencies are largely unchanged because most users have continued to egress via the LOS PoP since service's launch in March 2023.

\begin{figure}[!t]
    \includegraphics[width=\columnwidth]{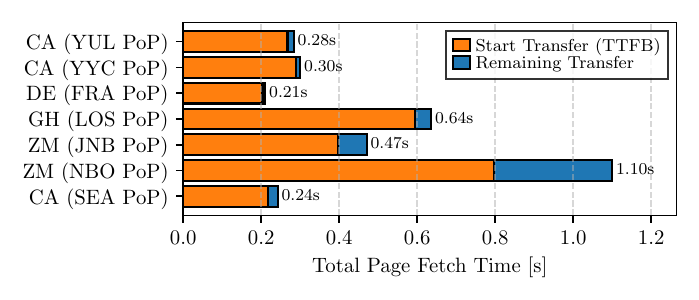}
    \vspace*{-3em}
    \caption{TTFB and total page fetch time to Cloudflare hosted websites from controlled nodes}
    \label{fig:total-page-fetch-time}
    \vspace*{-1em}
\end{figure}

\Cref{fig:total-page-fetch-time} presents the time-to-first-byte (TTFB) and total page fetch time from controlled vantage points to Cloudflare-hosted websites, which reflects the overall real-world browsing experience.
TTFBs are significantly lower for Starlink users in DE and CA due to rich edge-cache presence adjacent to their local PoPs.
The remaining transfer time until the page is fully fetched is relatively small, indicating that popular content is likely cached on nearby CDN servers.
In Ghana (GH), Starlink users experience significantly higher TTFB, with a median of total page fetch time of $\approx$0.64~s.
This is likely due to a combination of factors, including the relatively fewer CDN edge servers hosting popular content within West Africa, and extra hops to fetch content still served from outside the region.
We also highlight the impact of PoP reassignment from NBO to JNB in April 2025 from the Starlink dish in ZM.
TTFB decreased by $\approx$400~ms after the PoP change, and the attribution of remaining transfer time is also significantly reduced, reducing the total page fetch time to $\approx$0.5~s, which reflects the content concentration in JNB.
\begin{table}[!t]
    \centering
    \resizebox{\columnwidth}{!}{%
    \begin{tabular}{@{}ccccccc@{}}
    \hline
    \textbf{DE (FRA)} & \textbf{CA (SEA)} & \textbf{CA (YYC)} & \textbf{CA (YUL)} & \textbf{GH (LOS)} & \colorbox{LightRed}{\textbf{ZM (NBO)}} & \colorbox{LightBlue}{\textbf{ZM (JNB)}} \\ \hline
    95.03\%\textpm4.8\%    & 94.39\%\textpm3.3\%    & 97.55\%\textpm1.4\%    & 98.32\%\textpm1.2\%    & 85.09\%\textpm11.7\%        & 63.38\%\textpm7.6\%    & 92.95\%\textpm5.5\%        \\ \hline
    \end{tabular}%
    }
    \caption{Cache hit rate to Cloudflare-hosted websites from controlled nodes, defined as ``Country (PoP)''. \textcolor{red}{Red} denotes older PoP while \textcolor{blue}{Blue} shows newer PoP assignment.}
    \label{fig:cf-cache-hit}
    \vspace*{-2.5em}
    \end{table}
\Cref{fig:cf-cache-hit} shows Cloudflare cache-hit rates from the controlled vantage points.
For ZM, cache hits increased from $\approx$60\% to $\approx$90\% after NBO$\rightarrow$JNB reassignment, reducing origin fetches and stabilizing TTFB.
We also observe that LOS-adjacent vantage points can underperform when locally popular content is not well cached near LOS, which can force requests to more remote CDNs.

\subsection{Global PoP Reassignment}\label{sec:other_reassignments}

While African reassignments yielded dramatic improvements, PoP changes in infrastructure-rich regions reveal strikingly different dynamics.
In early 2025, Starlink activated Calgary (YYC) and Montréal (YUL) PoPs, transitioning Canadian users from US PoPs (Seattle (SEA), and New York City (NYC)) to domestic infrastructure.
Despite reducing geographic distance by 500--1500\,km, end-to-end CDN latencies remain largely unaffected, even paradoxically increasing by 4--5\,ms.
This counter-intuitive result stems from NA's mature CDN architecture: while SEA and NYC PoPs directly peer with massive CDN clusters, YYC and YUL require slightly longer terrestrial paths to reach the same cache servers or in Toronto (YYZ).
However, we observe increased cache hit rates (see~\Cref{fig:cf-cache-hit}) from Canadian clients mapped to in-country YYC and YUL PoPs compared to US-based SEA PoP.
This improvement reflects better alignment between user geography and CDN content localization as CA-based PoPs receive CA-specific content requests that match cached regional content.
\begin{figure}[t!]
    \begin{minipage}{0.65\linewidth}
        \centering
            \vspace{0.5em}
            \includegraphics[width=\columnwidth]{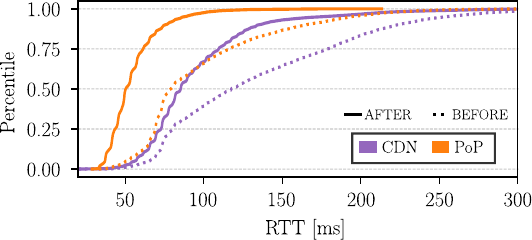}
            \label{fig:guam-cf-cdn-cdf-before-vs-after}
    \end{minipage}
    \begin{minipage}{0.26\linewidth}
            \centering
            \includegraphics[width=\columnwidth]{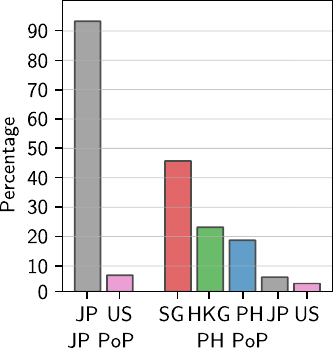}
        \label{fig:gu_cdn_servers_pop_reassignment}
    \end{minipage}
    \vspace*{-2.0em}
    \caption{CDN performance for Guam (GU) before and after the PoP reassignment from Japan to the Philippines. Left shows the CDF of CDN and PoP RTTs and Right shows the spread of CDN server locations.}
    \Description{Guam PoP reassignment}
    \label{fig:guam_pop_reassignment}
    \vspace*{-1.5em}
\end{figure}

Starlink infrastructure advancements in Asia-Pacific highlight a different perspective.
\Cref{fig:guam_pop_reassignment} shows how transitioning from Tokyo to Manila fundamentally reshaped web performance for Guam (GU) users.
Prior to the reassignment, 95\% of traffic was mapped to Japanese edges over consistent 2500\,km paths.
Post-reassignment, traffic scatters closer to PH: across SG (46\%), HK (23\%), PH (19\%), JP (6\%) and US (4\%).
This dispersion occurs because Manila's PoP peers with multiple regional IXPs, each offering different CDN presence.
SG and HK maintain extensive caches, but PH edges show 18\% higher miss rates, forcing origin fetches that inflate tail latencies.
Despite dispersion, we observe that PoP reassignment improved median RTT to PoP and CDN by 25\%, with a significant reduction in long tail.
Similar trends are also observed in Africa where PoP change from Germany to South Africa for Benin and Madagascar users resulted in more CDN location dispersion but with significant latency improvements (see \Cref{app:starlink-vs-terrestrial}).

\begin{tcolorbox}[title=\textit{Takeaway \#3}, enhanced, breakable]
PoP reassignments yield dramatically different benefits depending on regional CDN maturity.
African users experience 60\% latency reduction and improved cache hit rates when moving from European to continental PoPs, demonstrating that infrastructure placement (not satellite density) determines performance in under-served regions.
Conversely, reassignments in CDN-dense regions (CA) show marginal impact ($\pm$5\,ms) despite similar geographic distance reductions, as abundant local infrastructure makes PoP location largely irrelevant.
This differential impact reveals that optimal PoP placement must consider existing CDN topology: new PoPs transform performance in infrastructure-sparse regions but offer diminishing returns where CDN presence already saturates.
\end{tcolorbox}

%% file: sections/related.tex
\section{Related Work}\label{sec:related-works}

\noindent\textbf{CDN performance over TNs and LSNs}.
A rich body of work has dissected CDN request-mapping techniques and the performance over terrestrial networks~\cite{adhikari2014measurement,calder2015analyzing,saverimoutou2019influence,wohlfartLeveragingInterconnectionsPerformance2018,gigisSevenYearsLife2021}.
Since Starlink's rise, researchers have probed latency, loss, and throughput~\cite{michel2022,pan2023measuring,zhaoLENSLEOSatellite2024b,garcia2023multi} as well as the QoE of web browsing, live video streaming, video conferencing, and cloud gaming~\cite{izhikevich2023democratizing,kassem2022,multifacetedStarlink-www24,zhao2023real,zhaoLowLatencyLiveVideo2024b}.
Yet the compounded effect of content delivery components over LSNs, i.e., PoP assignment, DNS resolution, and CDN mapping on the perceived user experience has remained largely uncharted.

Several studies have attempted to characterize global CDN performance over Starlink, yet they often lack detailed insights from underserved regions such as Africa.
Studies have shown that African web traffic is often served from Europe or the United States~\cite{fanouExploringAnalysingAfrican2018}, and that sub-optimal DNS paths arise from poor peering~\cite{mbeweQoEImpactDoH2021a}.
\citet{leoCDN-hotnets2024} offered an initial look at HTTP metrics over Starlink in the region but did not separate the DNS resolver and CDN components of content delivery.
In addition, the authors reported significantly higher latencies for Starlink users in Africa compared to local terrestrial ISPs, since the users were associated with PoPs in Europe.
Starlink has since deployed new PoPs in Nairobi and Johannesburg in early 2025, alongside the existing sole African PoP in Lagos.
The evolution of Starlink's network infrastructure, particularly in underserved regions, and its impact on CDN performance, such as user-PoP assignment, DNS resolution, CDN mapping, and web performance, were largely unexplored in existing studies.

\smallskip
\noindent\textbf{Space-based content delivery architecture}.
Several proposals envision storing content and even computing on board LSNs or at their ground stations, thereby shortening the delivery path and smoothing mobility~\cite{lai2023starfront,spache2025,starcdn-sigcomm2025}.
\citet{lai2023starfront} proposed to use LEO satellites as ``cache in space'' resources and as low-latency space paths to terrestrial clouds, employing algorithms to judiciously place content replicas and assign user requests in a cost-effective manner to minimize operational expenses while meeting various latency requirements.
\citet{spache2025} proposed to integrate web cache directly into LEO satellites, utilizing communication schedules to manage prefetching and cache partitioning.
\citet{starcdn-sigcomm2025} addresses the LEO satellite mobility issues by introducing a consistent hashing scheme to reduce redundant content storage across dynamic satellite caches and a relayed fetching technique that counters the orbital motion by allowing cached content to effectively flow backward.
While these designs are compelling, their evaluations draw on limited testbeds and leave open questions about today's performance baseline.
We argue that a rigorous audit of current DNS and CDN behaviours over LSNs is therefore a prerequisite for an informed architectural change.

%% file: sections/discussion.tex
\section{Conclusion}\label{sec:discussion}

This paper demonstrated that LSN architecture fundamentally disrupts CDN locality assumptions, with performance impact varying significantly by regional infrastructure density.
While the satellite ``bent-pipe'' adds only 25--30\,ms, cascading mis-alignments through DNS and CDN layers compound latencies by up to 10$\times$.
The primary bottleneck is not satellite infrastructure but the mismatch between user location and PoP egress.
DNS-based CDNs select edges near resolvers rather than users, adding 40--100\,ms when resolvers are distant from PoPs.
Cache misses amplify these penalties: African users with 80\,ms resolver latency experience 250--300\,ms resolution times when recursive lookups are required.
These architectural mis-alignments explain why reassignments to local PoPs yield 60\% improvement in Africa but negligible change in Canada.
Surprisingly, Starlink outperforms terrestrial ISPs in regions with poor peering; for instance Benin and Madagascar terrestrial ISPs route to Europe via submarine cables (>7000\,km), while Starlink exits at continental PoPs, reducing paths by 30--60\,ms.
Our study provides constructive takeaways for both CDN and LSN operators.
For former, we highlight the need for satellite-specific request mapping that distinguishes between user location and network egress.
Current EDNS Client Subnet implementations and geolocation databases assume terrestrial topology, systematically mis-localizing satellite users.
For latter, co-locating PoPs with regional IXPs yields greater benefit than satellite density.
Moving forward, our efforts necessitates coordinated evolution of both LSN and CDN infrastructure.

%% file: sections/app-ethics.tex
\section{Ethics Statement} \label{app:ethics}

This research does not raise any ethical issues and no ethical approval is required. The data used in this study is publicly available and does not contain any personally identifiable information.

%% file: sections/app-measurement_config.tex
\section{CDN-Hosted Website Target List} \label{app:cdn_list}

\Cref{table:website-list} presents the web domains hosted by the three major CDN providers—Cloudflare, Akamai, and CloudFront—used in our active CDN and DNS measurements.

\begin{table*}[t]
  \centering
  \setlength{\tabcolsep}{6pt}  %
  \begin{tabularx}{\linewidth}{@{}l l X@{}}
    \toprule
    \textbf{CDN} & \textbf{Platform} & \textbf{Domains} \\ \midrule
    \multirow{2}{*}{Cloudflare}
      & RIPE Atlas + Controlled Nodes &
        \texttt{www.broadcom.com}, \texttt{www.comodoca.com},
        \texttt{www.epicgames.com}, \texttt{www.apnic.net},
        \texttt{www.riskified.com}, \texttt{www.wiley.com},
        \texttt{www.vmware.com}, \texttt{www.sportskeeda.com},
        \texttt{www.garmin.com}, \texttt{www.fao.org},
        \texttt{www.n-able.com}, \texttt{www.linkedin.com} \\ \cmidrule(l){2-3}
      & Controlled Nodes &
        \texttt{www.roku.com}, \texttt{www.sourceforge.net},
        \texttt{www.namecheap.com}, \texttt{www.openai.com},
        \texttt{www.cpanel.net}, \texttt{www.zendesk.com},
        \texttt{www.17track.net}, \texttt{www.quora.com},
        \texttt{www.temu.com}, \texttt{www.constantcontact.com},
        \texttt{www.fanfiction.net}, \texttt{www.fao.org},
        \texttt{www.matterport.com}, \texttt{www.techtarget.com} \\ \midrule
    Akamai & RIPE Atlas + Controlled Nodes &
        \texttt{www.microsoft.com}, \texttt{www.apple.com},
        \texttt{www.bing.com}, \texttt{www.icloud.com},
        \texttt{www.intuit.com}, \texttt{www.unity3d.com},
        \texttt{www.samsung.com}, \texttt{www.ebay.com},
        \texttt{www.webex.com}, \texttt{www.cisco.com} \\ \midrule
    CloudFront & RIPE Atlas + Controlled Nodes &
        \texttt{www.soundcloud.com}, \texttt{www.zynga.com},
        \texttt{www.doi.org}, \texttt{www.booking.com},
        \texttt{www.brave.com}, \texttt{www.tycsports.com},
        \texttt{www.logitech.com}, \texttt{www.checkpoint.com},
        \texttt{www.goodreads.com}, \texttt{www.surveymonkey.com} \\
    \bottomrule
  \end{tabularx}
  \caption{Targeted website domains, their CDN hosting operator and the measurement platform used. \label{table:website-list}}
\end{table*}

%% file: sections/app-additional_analysis.tex
\section{Additional Analysis} \label{app:additional}

In this section, we present the additional analysis to support the takeaways in the paper.

\subsection{Cloudflare CDN Performance Across Africa} \label{app:africa-cloudflare-aim}

\begin{figure}[tb]
    \includegraphics[width=\columnwidth]{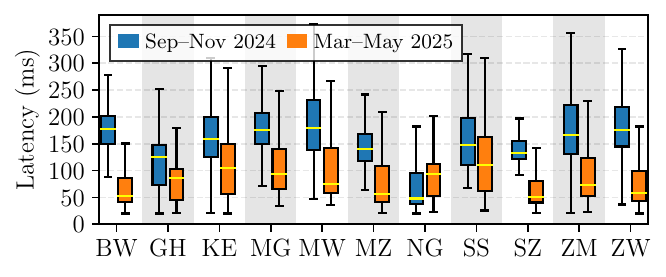}
    \caption{Latency to the ``best'' performing Cloudflare CDN server locations from African countries before and after the PoP reassignments in early 2025.}
    \label{fig:cdf_af_cloudflare_all_countries}
  \end{figure}

\Cref{fig:cdf_af_cloudflare_all_countries} presents a comparison between baseline latency to the ``best'' performing Cloudflare CDN server (extracted from the Cloudflare AIM dataset) across African countries over a three-month period before and after the African PoP reassignments in early 2025.
Nearly all countries demonstrate approximately 2$\times$ improvements in latency, with latency levels in May 2025 reaching those observed in South America, the USA, and Europe.
As discussed in~\Cref{sec:africa_case}, this enhancement primarily benefits from Starlink's PoP reassignments in early 2025, which directed African users to one of the African PoPs in Nairobi (KE), Lagos (NG), or Johannesburg (ZA).
As a result, CDN traffic are more likely routed to local CDN servers within Africa, significantly reducing the latency.

\subsection{DNS Resolution Times} \label{app:dns-resolution-latency}

\begin{figure}[tb]
  \includegraphics[width=\columnwidth]{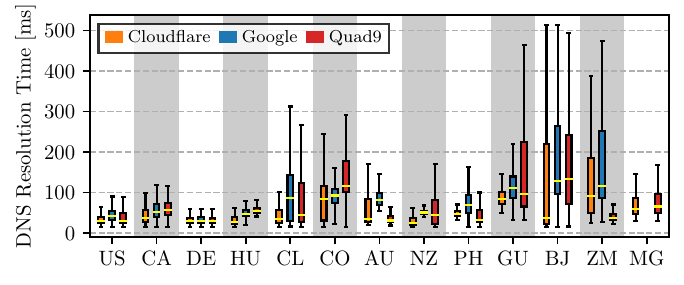}
  \caption{DNS resolution times across three major public resolvers for websites in our CDN website list.}
  \label{fig:dns_resolution_times_three_providers_appendix}
  \Description{DNS resolution times across different public resolvers.}
\end{figure}

\Cref{fig:dns_resolution_times_three_providers_appendix} presents a comparison of DNS resolution times for domain names associated with the websites from our curated CDN-hosted website list within the top 500 in the Tranco~\cite{Tranco} list.
Our analysis reveals that South American (CO and CL) \& African (BJ and ZM) countries experience higher resolution latencies compared to other regions.
The users of countries, whose home PoPs are  situated in Nairobi (KE) and Lagos (NG) are particularly affected.
Similarly, South American countries like Chile and Colombia, which connect to local PoPs, also exhibit marginally increased resolution delays.
This phenomenon may stem from additional recursive DNS lookups, as domain names are likely not cached in resolvers serving these regions.
The latency challenges for these countries are further compounded by delays in reaching the DNS resolvers themselves.

\subsection{Suboptimal DNS Resolver Selection} \label{app:distant_dns_mapping}

\begin{figure}[t]
  \includegraphics[width=\columnwidth]{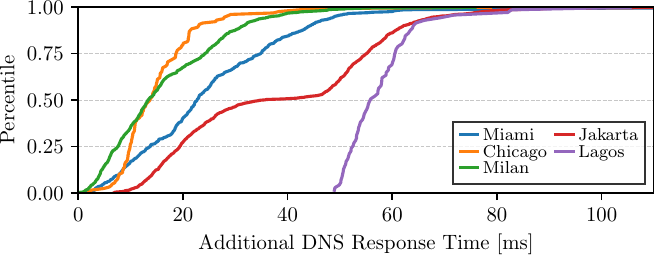}
  \caption{Additional DNS response time for Cloudflare and Google DNS resolvers when mapped to DNS resolvers further away from these Starlink PoP locations (i.e., DNS Response Time $-$ PoP latency)}
  \label{fig:dns_additional_latency}
\end{figure}

On multiple occasions, we have observed distant DNS resolver mappings from Starlink clients associated with several PoPs in different regions.
As shown in~\Cref{fig:dns_additional_latency}, when DNS requests are routed to distant resolvers that are not co-located with the Starlink PoP, it can increase DNS response times, and more critically, DNS-based CDNs may map users to suboptimal remote CDN servers, leading to increased latency when accessing content.
This could happen due to the internal load balancing policies of these public DNS providers, or anycast routing fluctuations.
For example, Starlink users served by the Jakarta (ID) PoP were consistently routed to a Google DNS resolver instance in Singapore (>900 km away from the Jakarta PoP).
Starlink users associated with the Milan PoP can be directed to Google and Cloudflare DNS resolver instances in nearby European cities (Zurich, Marseille, Amsterdam, and Sofia).
The suboptimal DNS resolver selection is more pronounced for Starlink users in regions with less developed terrestrial infrastructure.
In well-connected regions, the impact is less significant even when internal load balancing happens, due to the availability of multiple nearby DNS resolver instances.

\begin{figure}[t]
    \includegraphics[width=\columnwidth]{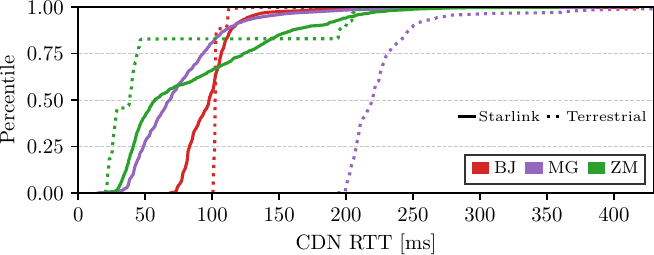}
    \caption{\label{fig:africa_cdn_server_stl_vs_terr} Starlink vs Terrestrial: Latency to CDN servers for Akamai hosted websites for BJ, MG and ZM}
\end{figure}

\begin{figure}[t]
    \begin{subfigure}[h]{0.35\columnwidth}
        \centering
        \includegraphics[height=1.5in]{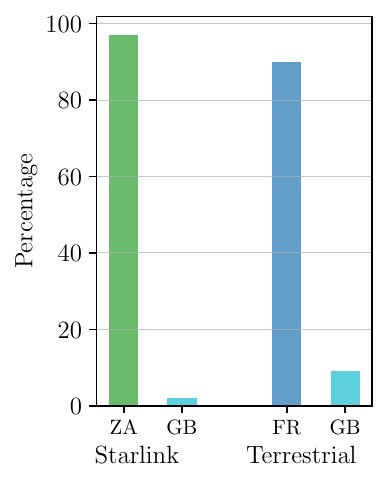}
        \caption{\label{fig:bj_apple_cdn_server_stl_vs_terr}Benin}
    \end{subfigure}
    \begin{subfigure}[h]{0.60\columnwidth}
        \centering
        \includegraphics[height=1.5in]{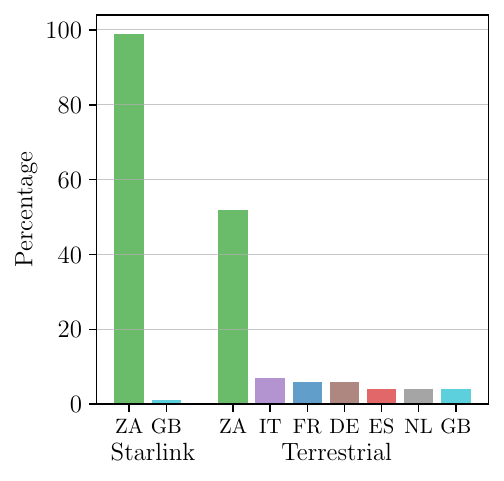}
        \caption{\label{fig:mg_apple_cdn_server_stl_vs_terr}Madagascar}
    \end{subfigure}
    \caption{\label{fig:resolved_ip_geolocation_BJ_and_MG}Resolved IP geolocation for Akamai hosted sites for Starlink and Terrestrial ISP users in BJ and MG, indicating the number of times connecting to a CDN server location in a particular country}
\end{figure}

\subsection{Africa: Starlink vs Terrestrial} \label{app:starlink-vs-terrestrial}

Starlink's popularity as a global ISP is emerging in Africa in recent years, as terrestrial fiber infrastructure is still under-developed in the region.
We depict the Starlink and terrestrial ISP users latency distribution to content hosted by Akamai CDN in~\Cref{fig:africa_cdn_server_stl_vs_terr}.
While Zambia (ZM) terrestrial users are routed to South Africa (a content rich region), because of its proximity, terrestrial users in other African nations, like Benin (BJ) and Madagascar (MG) are mostly routed to CDN servers in Europe, observing high latency-to-content.
However, newer Starlink PoPs in Nairobi and Johannesburg allow users from these countries to fetch most content within the continent, observing significantly lower latencies bypassing terrestrial connections.
For example, Madagascar Starlink users observe around 150~ms lower median latency than their terrestrial counterparts.
Meanwhile, Starlink users in Zambia experience slightly higher median latencies, a delta of 20~ms, largely due to Starlink's ``bent-pipe'' connection.
The differences for Starlink users in Benin (connecting to the Lagos PoP) and terrestrial users are more subtle.
In around 50\% of the cases, Starlink provides a lower latency-to-content (connects to South African CDN servers) than terrestrial ISP users.
\Cref{fig:resolved_ip_geolocation_BJ_and_MG} depicts the country where the CDN server was located when connecting to Akamai-hosted websites from Starlink and TN users in Benin and Madagascar.
It is interesting to note that all Starlink PoPs in Africa so far are in English-speaking countries.